\documentclass[aps,twocolumn,prl,amsmath,amssymb,superscriptaddress,floats]{revtex4}

\usepackage{graphics}
\usepackage{hyperref}
\usepackage{epsfig}
\usepackage{color}
\usepackage{multirow}
\usepackage{bm}
\usepackage{float}

\newcommand{\upperRomannumeral}[1]{\uppercase\expandafter{\romannumeral#1}}

\begin{document}

\title{
Antiferromagnetic resonance and terahertz continuum in $\alpha-$RuCl$_3$}


\author{A. Little}
\affiliation{Department of Physics, University of California, Berkeley CA 94720, USA}
\affiliation{Materials Science Division, Lawrence Berkeley National Laboratory, Berkeley CA 94720, USA}

\author{Liang Wu}
\email{liangwu@berkeley.edu}
\affiliation{Department of Physics, University of California, Berkeley CA 94720, USA}
\affiliation{Materials Science Division, Lawrence Berkeley National Laboratory, Berkeley CA 94720, USA}
\affiliation{Department of Physics and Astronomy, University of Pennsylvania, Philadelphia, Pennsylvania 19104, USA}

\author{P. Lampen-Kelley}
\affiliation{Department of Materials Science and Engineering, University of Tennessee, Knoxville, TN 37996, U.S.A.}
\affiliation{Materials Science and Technology Division, Oak Ridge National Laboratory, Oak Ridge, TN, 37831, U.S.A.}

\author{A. Banerjee}
\affiliation{Quantum Condensed Matter Division, Oak Ridge National Laboratory, Oak Ridge, Tennessee 37830, USA.}

\author{S. Patankar}
\affiliation{Department of Physics, University of California, Berkeley CA 94720, USA}
\affiliation{Materials Science Division, Lawrence Berkeley National Laboratory, Berkeley CA 94720, USA}

\author{D. Rees}
\affiliation{Department of Physics, University of California, Berkeley CA 94720, USA}
\affiliation{Materials Science Division, Lawrence Berkeley National Laboratory, Berkeley CA 94720, USA}

\author{C. A. Bridges}
\affiliation{Chemical Sciences Division, Oak Ridge National Laboratory, Oak Ridge, Tennessee 37830, USA.}

\author{J.-Q. Yan}
\affiliation{Material Sciences and Technology Division, Oak Ridge National Laboratory, Oak Ridge,Tennessee 37830, USA.}

\author{D. Mandrus}
\affiliation{Department of Materials Science and Engineering, University of Tennessee, Knoxville, TN 37996, U.S.A.}
\affiliation{Materials Science and Technology Division, Oak Ridge National Laboratory, Oak Ridge, TN, 37831, U.S.A.}

\author{S. E. Nagler}
\affiliation{Quantum Condensed Matter Division, Oak Ridge National Laboratory, Oak Ridge, Tennessee 37830, USA.}
\affiliation{Bredesen Center, University of Tennessee, Knoxville,Tennessee 37966, USA.}

\author{J. Orenstein}
\affiliation{Department of Physics, University of California, Berkeley CA 94720, USA}
\affiliation{Materials Science Division, Lawrence Berkeley National Laboratory, Berkeley CA 94720, USA}

\date{\today}

\begin{abstract}
 We report measurements of optical absorption in the zig-zag antiferromagnet $\alpha$-RuCl$_3$ as a function of temperature, $T$, magnetic field, $B$, and photon energy, $\hbar\omega$ in the range $\sim$ 0.3 to 8.3 meV, using time-domain terahertz spectroscopy. Polarized measurements show that 3-fold rotational symmetry is broken in the honeycomb plane from 2 K to 300 K. We find a sharp absorption peak at 2.56 meV upon cooling below the N\'{e}el temperature of 7 K at $B=0$ that we identify as magnetic-dipole excitation of a zero-wavevector magnon, or antiferromagnetic resonance (AFMR).  With application of $B$, the AFMR broadens and shifts to lower frequency as long-range magnetic order is lost in a manner consistent with transitioning to a spin-disordered phase. From direct, internally calibrated measurement of the AFMR spectral weight, we place an upper bound on the contribution to the $dc$ susceptibility from a magnetic excitation continuum.

\end{abstract}

\maketitle

When exchange interactions between neighboring spins in a magnetic system are at odds, the resulting frustration can lead to a highly entangled form of matter with no ordered ground state. Such highly correlated, liquid-like states have come to be known as quantum spin liquids (QSLs) \cite{balents2010spin, savary2016quantum}. The QSL state is markedly featureless and difficult to experimentally detect -- there being no local order parameter or phase transition. Nonetheless QSL candidates are of great interest both theoretically and experimentally because they can host emergent fractionalized excitations -- wherein the electron is divided into quasiparticles with fractional quantum numbers \cite{sachdev2008quantum}.

Lattices exhibiting geometric frustration, specifically those based on triangular arrangements of spins such as the Kagome \cite{han2012fractionalized}, have long been at the center of QSL research. A significant step in the development of QSL theory was an alternative, exactly solvable route proposed by Kitaev~\cite{kitaev2003fault, kitaev2006anyons}. The Kitaev spin liquid (KSL) model consists of spin-1/2 particles arranged on a hexagonal lattice with Ising exchange interaction between nearest neighbors. Frustration results from rotation of the Ising axis from bond to bond, rather than the geometry of the lattice. In the exact solution of the KSL model the spin Hamiltonian is recast in terms of Majorana fermions propagating on the landscape of a static Z$_2$ gauge field \cite{kitaev2006anyons}. Exact analytical results for dynamical spin correlations can be derived \cite{baskaran2007exact}, leading to predictions for the signatures of Majorana quasiparticles in inelastic neutron \cite{knolle2014dynamics,knolle2015dynamics,song2016low}, Raman \cite{knolle2014raman}, and resonant X-ray \cite{halasz2016resonant} scattering.

Coupled with theoretical progress, interest in the KSL model was greatly stimulated by the suggestion \cite{jackeli2009mott, chaloupka2010kitaev} that Kitaev interactions could arise in real materials, such as iridates and ruthenates \cite{williams2016incommensurate,modic2014realization, plumb2014alpha}, as a natural consequence of spin-orbit coupling. Although it was found that these materials order magnetically at low $T$ \cite{liu2011long, choi2012spin, chun2015direct, biffin2014noncoplanar, takayama2015hyperhoneycomb, kim2015kitaev, sears2015magnetic, cao2016low} interest in these systems as proximate Kitaev spin liquids has developed, accelerated by the idea that emergent KSL quasiparticles may exist despite the presence of magnetic order. $\alpha$-RuCl$_3$ has risen to prominence in this line of research because crystals suitable for inelastic neutron scattering (INS) have been grown, whereas INS is notoriously difficult in iridate compounds. INS performed on $\alpha$-RuCl$_{3}$ indicates a continuum of excitations extending to 15 meV and centered at zero in-plane wavevector, in addition to magnon peaks below the N\'eel temperature, $T_N$ \cite{banerjee2016proximate, banerjee2016neutron}. This spectrum has been interpreted in terms of the $\mathbf{q}=0$ dynamical susceptibility of KSLs, in which fractionalization into Majorana fermions and $Z_2$ vortices creates a continuum of spin fluctuations above a small gap \cite{knolle2014dynamics, knolle2015dynamics,song2016low}. Interpretations in terms of an incoherent multi-magnon continuum have also been advanced \cite{winter2017breakdown}. The search for spin liquid states in $\alpha$-RuCl$_3$ has been further stimulated by the observation that magnetic order is destroyed by in-plane magnetic fields that are weak compared to the leading order exchange interactions, suggesting the existence of one or more quantum critical points and a variety of exotic phases occupying the $B-T$ phase space \cite{yadav2016kitaev, sears2017phase, baek2017observation, hentrich2017large,  zheng2017gapless, leahy2017anomalous}.

Thus far, the dynamical response in $\alpha$-RuCl$_3$ has been probed exclusively by inelastic scattering \cite{sandilands2015scattering, banerjee2016proximate, banerjee2016neutron, ran2017spin, do2017incarnation}. In this work, we use time-domain THz spectroscopy (TDS) to probe excitations in $\alpha$-RuCl$_{3}$ in the frequency range 0.08-2 THz (energy range 0.3-8.3 meV) and magnetic field range 0-7 Tesla. TDS is complementary to INS in exploring magnetic excitations, as it focuses on the $\textbf{q}=0$ response function with higher spectral resolution and precise, internally calibrated, determination of absolute spectral weight. By contrast, INS accesses near zero in-plane wavevector ($\textbf{q}_{ab}$) by selecting non-zero out-of-plane momenta ($\textbf{q}_c$), introducing broadening and distortion of lineshapes from dispersion along the $c$-direction. Furthermore, INS studies of $\alpha$-RuCl$_3$ at $\textbf{q}_{ab}=0$ published to date are limited to energies above $\sim$ 2 meV by the elastic scattering background. The ability of TDS to trace the spectrum and spectral weight of the magnetic response function to lower energies at high resolution is critical for achieving a theoretical understanding the effective spin Hamiltonian of $\alpha$-RuCl$_3$ and the nature of its phases in the $B-T$ plane.

The crystals used in this study exhibit a single thermal phase transition to zig-zag antiferromagnetic order at a $T_N$ $\sim$ 7 K and have been shown to contain few stacking faults \cite{banerjee2016neutron}. Samples of $\alpha$-RuCl$_3$ with typical area $\sim 0.8$ cm$^2$ and thickness 1 mm were mounted over an aperture on a copper plate. We measured THz transmission at near normal incidence such that the probing fields lie in the $ab$ (honeycomb) plane.

TDS is based on measuring the transmission coefficient, $t(\omega)$, of picosecond timescale electromagnetic pulses. In the weak absorption limit appropriate to a large gap Mott insulator such as $\alpha$-RuCl$_{3}$, $|t(\omega)|\cong[4n/(n+1)^2]\exp{[-\alpha(\omega)d]}$, where $\alpha(\omega)$ is the frequency dependent absorption coefficient, $n$ is the index of refraction, and $d$ is the sample thickness (See Supplementary Information (SI) section \upperRomannumeral{1}  \cite{SI}).

Before considering the frequency-dependence of the absorption, we show that TDS probes the point group symmetry of the unit cell of  $\alpha$-RuCl$_3$. In the presence of 3-fold rotational symmetry ($C_3$), $t(\omega)$ will be independent of the direction of the THz field in the $ab$ plane. To test for $C_3$, we measured $t(\omega)$ as the sample was rotated between a pair of crossed linear polarizers. The inset to Fig. 1a shows a polar plot of the transmitted amplitude as a function of sample angle at room temperature. The observed anisotropy demonstrates that $C_3$ is broken at 300 K. The 4-fold pattern of the polar plot indicates optical birefringence, that is the existence of a pair of orthogonal principal axes with distinct values of the index of refraction.  Laue X-ray diffraction on the same crystal confirmed that these directions correspond to the $a$ and $b$ axes depicted in Fig. 1b (see SI section \upperRomannumeral{2}  \cite{SI}). The optical birefringence is likely related to an in-plane distortion of the Ru hexagons in which the length of the pair of opposing Ru-Ru links parallel to the $b$ axis is greater than the other two by $\sim0.2\%$ \cite{johnson2015monoclinic,  ziatdinov2016atomic}.  Although there are three equivalent orientations of this distortion, we note that the crystal under study must comprise largely a single such domain on the scale of the optical probe ($\sim$5 mm$^2$ area by 1 mm thickness) in order to show strong optical anisotropy.

\begin{figure}[htp]
\centering
\includegraphics[width=1\columnwidth]{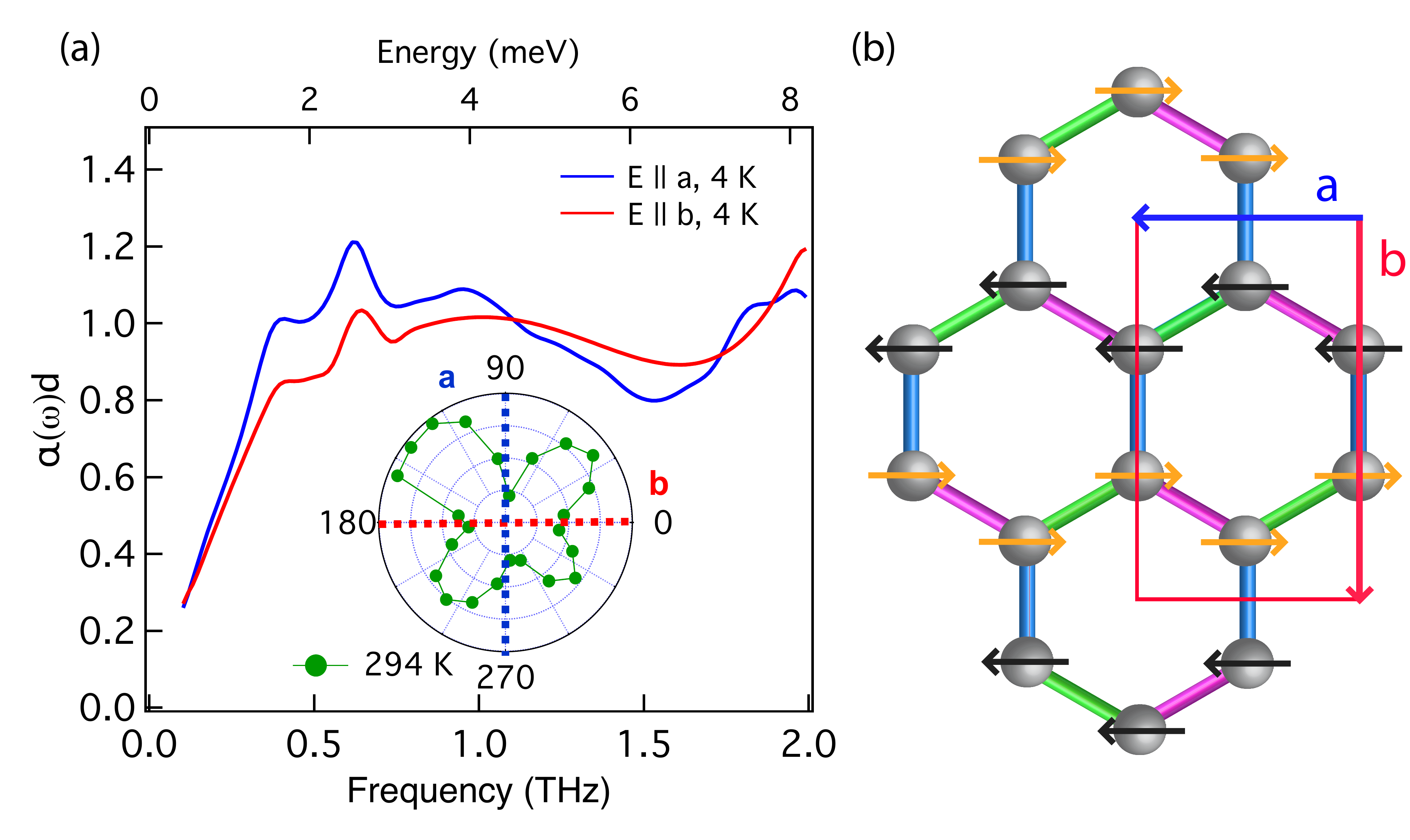}
\caption{(a) Optical density $\alpha(\omega)d$ with $\mathbf{E}$ parallel to axis $\textbf{a}$ (blue) and axis $\textbf{b}$ (red) at T = 4 K. Inset: Polar plot of transmitted THz electric field amplitude at 294 K as a function the rotation angle of a sample positioned between crossed polarizers. The principal axes are marked by dashed lines. (b) Zig-zag AFM order on the honeycomb lattice, with $\textbf{a}$ and $\textbf{b}$ axis directions denoted by blue and red arrows.}
\label{fig1}
\end{figure}

In the main panel of Fig. 1a, we plot the absorption $\alpha(\omega)d$ at 4 K with the $\mathbf{E}$ field polarized parallel to the $\textbf{a}$ and $\textbf{b}$ directions. A conspicuous feature of both spectra is the narrow peak at 0.62 THz (2.56 meV), which is superposed on a broad continuum of absorption with a low-energy cut-off. The spectra for the two orthogonal polarizations are distinctly different, showing that the breaking of $C_{3}$ observed at room temperature persists to low $T$. Thus the phase transition at 150 K (which we observe optically, see SI section \upperRomannumeral{3} \cite{SI}) must occur between crystal structures that each break $C_3$, for example monoclinic $\rightarrow$ triclinic~\cite{note_structure}.

Figs. 2a and 2b focus on the temperature dependence of the sharp peak in zero magnetic field. The inset to Fig. 2a compares pulses transmitted through the sample at 2 K and 15 K. In the main part of Fig. 2a we show, on an expanded vertical scale, the results of subtracting the THz transient measured at 15 K from those measured at various temperatures below the magnetic transition, for $\mathbf{B}(t)\perp\textbf{a}$. The oscillations that grow with decreasing $T$ are well described by damped sine waves $Ae^{-\Gamma t}\textrm{sin}(\omega_R t)$, where $A$ is the amplitude, $\omega_R$ is the resonant frequency, and $\Gamma$ is the decay rate (see SI section \upperRomannumeral{4} \cite{SI}.  Fig. 2b illustrates the $T$-dependence of $A$ (left-hand scale) and $\Gamma$ (right-hand scale).

\begin{figure*}[htp]
\centering
\includegraphics[width=0.87\textwidth]{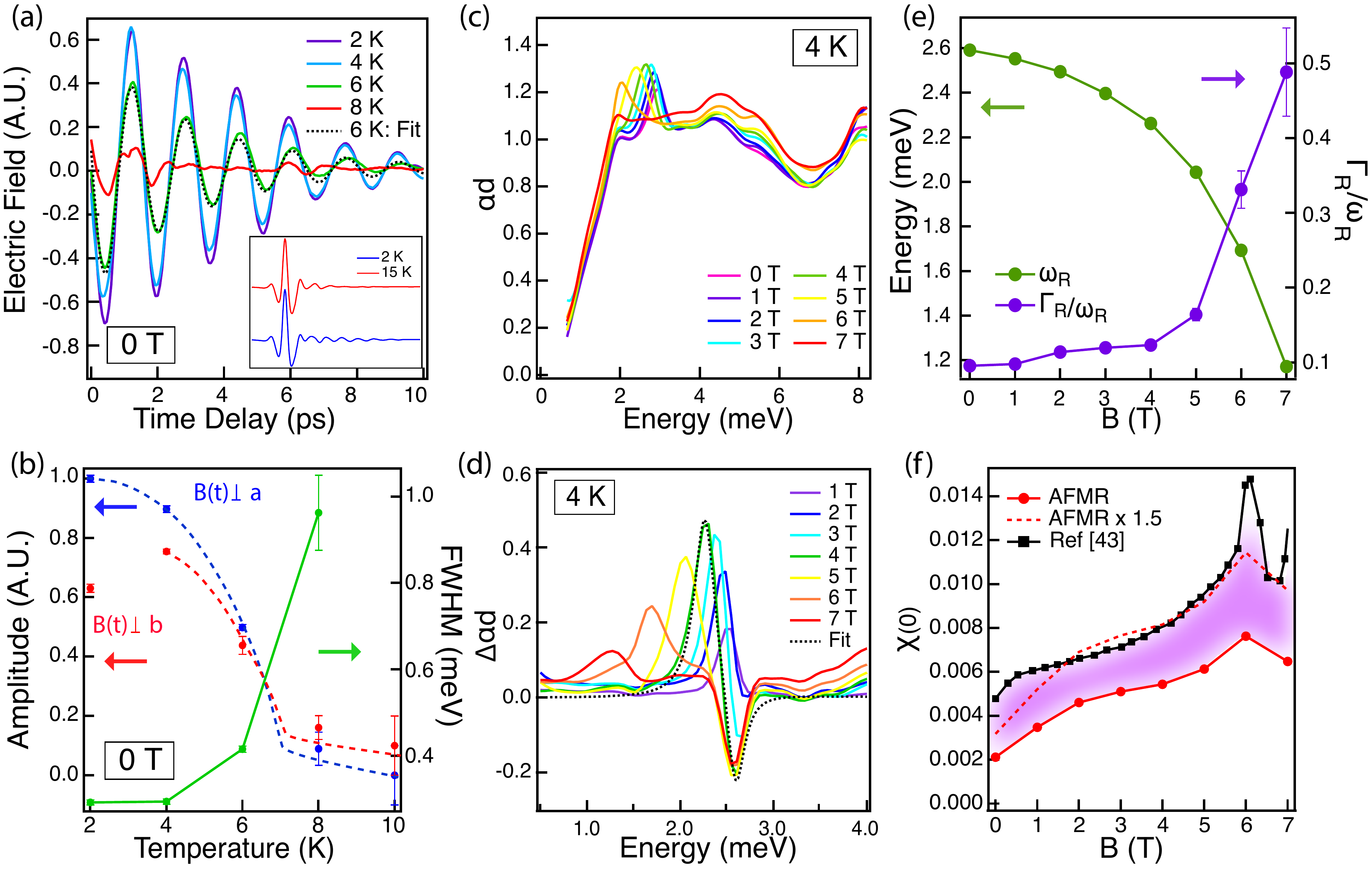}
\caption{(a) Coherent magnon emission measured in the time domain at 2 K, 4 K,  6 K, and 8 K on an expanded vertical scale. Inset: Time trace of transmitted THz $\mathbf{E}$ field at 2 K (blue) and 15 K (red). The 2 K pulse shows coherent magnon radiation while the 15 K pulse does not.  (b) Resonance amplitude (left-hand scale) with \textbf{B(t)}$\perp$\textbf{a} (blue) and  \textbf{B(t)}$\perp$\textbf{b} (red) and full-width half maximum (FWHM) along \textbf{a} (right-hand scale) as a function of temperature. Dashed lines are a guide to the eye. (c) The absorption spectrum at 4 K as a function of magnetic field.  (d) Absorption spectra with DC $B$- field parallel to THz field \textbf{B(t)}, both at 45$^{\circ}$ between the $a$ and $b$ axes. The zero-field spectrum is subtracted.  (e) Dependence of AFMR energy (left-hand axis) and inverse quality factor, $\Gamma_R/\omega_R$ (right-hand axis), on magnetic field. (f) Solid black and red circles show the static magnetic susceptibility, $\chi(0)$, and the contribution to $\chi(0)$ from the $\textbf{q}=0$ spin wave, respectively, as a function of magnetic field. The shaded region between indicates the maximum contribution from a magnetic excitation continuum.}
\label{fig2}
\end{figure*}

As the 2.56 meV mode appears at $T_N$, it is natural to associate it with resonant magnetic-dipole excitation of a $\textbf{q}=0$ magnon, which is known as antiferromagnetic resonance (AFMR)~\cite{keffer1952theory,nagamiya1951theory}. AFMR will appear at a nonzero frequency whenever $SU(2)$ spin rotation symmetry is broken by spin-orbit interactions, as are clearly present in $\alpha$-RuCl$_3$. However, as translational symmetry is changed at $T_N$, it is also conceivable that the resonance results from folding to zero wavevector of an acoustic phonon.

To test whether the resonance is indeed AFMR, we performed TDS as function of in-plane magnetic field from 0 to 7 Tesla, obtaining the absorption spectra shown in Fig. 2c. The resonant mode clearly shifts systematically to lower frequency with increasing $B$. As the periodicity of the antiferromagnetic order does not change with field~\cite{sears2017phase}, this observation demonstrates that the mode is not a zone-folded phonon and confirms its identity as AFMR .

Assuming that photons couple to the AFMR through the magnetic dipole interaction, we can evaluate the imaginary part  of the dynamic magnetic susceptibility at zero wavevector, $\chi_2(\omega)$, associated with the peak. To focus on this component we subtract the zero-field spectrum from those measured with $\textbf{B}\neq0$; the resulting difference spectra are shown in Fig. 2d. The strength of the absorption thereby is directly related to $\chi_2(\omega)$ via the relation,
\begin{equation}
\alpha(\omega)d=\frac{\omega nd}{2c}\chi_2(\omega)=\frac{\omega T_{rt}}{4}\chi_2(\omega)
\end{equation}
Note that the absolute, as opposed to relative, values of $\chi_2(\omega)$ are obtained directly from fundamental observables: optical density, $\alpha d$, and the pulse roundtrip time, $T_{rt}$ (see SI section \upperRomannumeral{1} \cite{SI}).

We find that for all values of the magnetic field the resonance can be well fit by a Lorentzian lineshape, that is,
\begin{equation}
\chi_2(\omega,B)=\frac {S\omega\Gamma}{(\omega^2-\omega_R^2)^2+\omega^2\Gamma^2},
\end{equation}
where $\omega_R$, $\Gamma$ are now field-dependent and $S(B)$ parameterizes the the overall amplitude. The dashed line in Fig. 2d illustrates the quality of the fit for the 4 Tesla difference spectrum (equally good fits for other fields are shown in SI section \upperRomannumeral{4} \cite{SI}). The variation with $B$ of the resonant frequency and inverse quality factor, $\Gamma/\omega_R$, obtained from the lineshape analysis are shown in Fig. 2e. The width of the resonance measured at zero applied field, $\approx$ 300 $\mu$eV, is at least 5 times smaller than the $\bf{q}_{ab}\approx 0$ peak observed by INS~\cite{banerjee2016neutron, banerjee2017excitations}. It is striking that although $\omega_R(B)$ decreases with increasing $B$, the resonance remains remains a well-defined, underdamped mode despite the loss of long-range magnetic order that occurs at a critical field, $B_c\approx$ 7 Tesla. Recent experiments that extend electron spin resonance measurements to higher fields show that this mode persists through the transition spin-disordered state; its frequency reaches a minimum value of $\approx$ 1 meV at $B_c$~\cite{ponomaryov2017direct} and thereafter increases linearly in proportion to $B-B_c$~\cite{ponomaryov2017direct, wang2017magnetic}.

A key issue in unravelling the physics of $\alpha$-RuCl$_3$, in particular its proximity to a spin liquid ground state, is the existence and strength of a continuum of magnetic excitations at $\textbf{q}=0$ in addition to well-defined magnon modes. THz spectroscopy directly addresses this issue by providing an auto-calibrated measurement of $\chi_2(\omega)$ at zero wavevector. The thermodynamic sum rule, derived from the Kramers-Kronig relation, relates $\chi_2(\omega)$ to the $dc$ magnetic susceptibility, $\chi(0)$,
\begin{equation}
\chi(0)=\frac{2}{\pi}\int_0^\infty\frac{\chi_{2}(\omega^\prime)}{\omega^\prime}d\omega^\prime.
\end{equation}
While Eq. 3 is valid in general, the contribution to the $dc$ susceptibility of a mode described by the Lorentzian lineshape of Eq. 2 is simply given by $\chi(0)=S/\omega_R^2$.

The thermodynamic sum rule allows us to place a bound on the strength of the $\textbf{q}=0$ magnetic continuum in $\alpha$-RuCl$_3$. In Fig. 2f we compare the $dc$ susceptibility associated with the spin wave resonance, $S(B)/\omega_R^2(B)$, with recent measurements of $\chi(0,B)$ using low-frequency susceptometry~\cite{banerjee2017excitations}. Both the spin wave contribution and the total $\chi(0,B)$ grow with increasing field, maintaining a fixed proportionality for $B<6$ T; this is highlighted by the dashed line, which shows $S(B)/\omega_R^2(B)$ scaled by a factor of 1.5. The shaded region between the two curves corresponds the $dc$ susceptibility not accounted for by the AFMR resonance. It is expected that in a quantum phase transition from a magnetically ordered phase to a QSL with fractional excitations, the spectral weight of spin wave modes would shift to a broadband magnetic continuum. Our spectra show instead that the contribution to the $dc$ susceptibility from a magnetic continuum remains comparable in size to the contribution of the $q=0$ spin wave, which remains a well-defined mode even approaching the critical magnetic field. This suggests that the $B_c\approx$ 7 T transition cannot be straightforwardly interpreted as a transition to a QSL.

Finally, we discuss the broad-band component of the THz absorption that is evident in Figs. 1a and 2c.  First, the thermodynamic sum rule argument described above rules out the possibility that the large observed continuum arises entirely from magnetic-dipole absorption. To show this, consider converting the entire $\alpha(\omega)$ to $\chi_2(\omega)$ using Eq. 1, and then integrating $\chi_2(\omega)/\omega $ with respect to $\omega$ to obtain a value for $\chi(0)$. As is already evident from comparison of the spectral weight of the resonant and broadband contributions to $\alpha d$, the $\chi(0)$ that emerges from this calculation is far larger, by $\approx$ 30 times, than the measured value of 0.02 emu/mole ($\sim$ 0.005 in SI units)~\cite{sears2015magnetic}. We conclude that the dominant contribution to the broadband absorption must originate from electric, rather than magnetic-dipole coupling, as expressed for example in terms of an optical conductivity.

Fig. 3 shows optical conductivity, $\sigma_1(\omega)$, at temperatures from 2 K to room temperature, converted from the absorption coefficient using the relation, $\sigma_1(\omega)=2nY_0\alpha(\omega)$, where $Y_0=377 \Omega^{-1}$ is the admittance of free space (see SI section I). A striking feature of the spectra is the lack of temperature dependence -- in particular the drop-off in $\sigma_1(\omega)$ below $\approx$ 1 meV remains well-defined even at high temperatures where $k_BT\gg$ 1 meV. The linear in $\omega$ cut-off below 1 meV evident in Figs. 1, 2c, and 3 is a highly reproducible feature seen in all spectra. Further evidence for the decrease in $\sigma_1(\omega)$ below 1 meV is that the $dc$ conductivity, $\sigma(0)$, (shown as a solid red circle) is indistinguishable from the origin on the scale of Fig. 3 even at room temperature, where $\sigma(0)\sim 3\times 10^{-4}$ $\Omega^{-1}$ cm$^{-1}$.

\begin{figure}[htp]
\centering
\includegraphics[width=0.7\columnwidth]{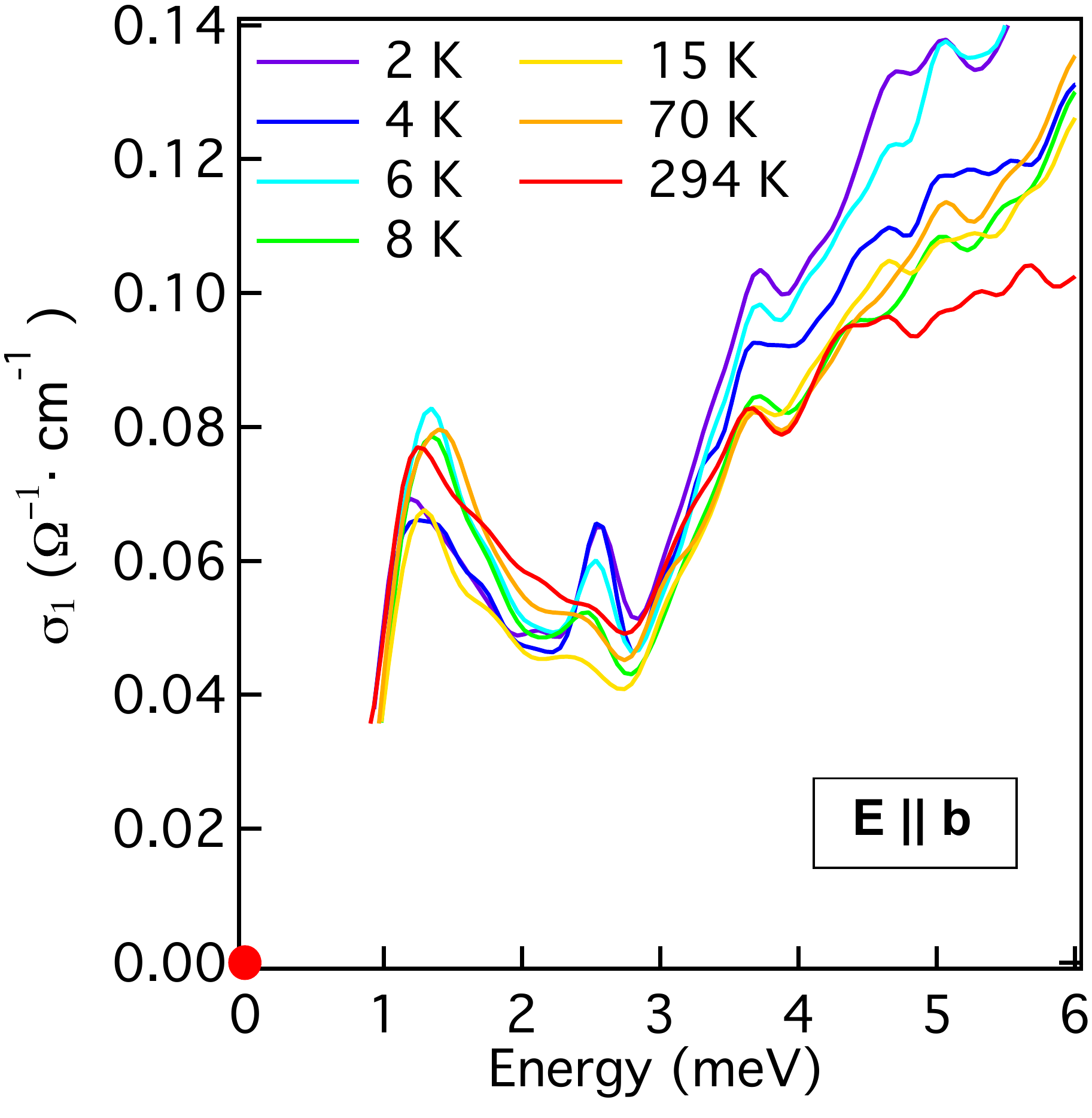}
\caption{Absorption spectra interpreted as optical conductivity, with \textbf{E} parallel to \textbf{b}.}
\label{fig3}
\end{figure}

At this point the origin of the broadband THz conductivity in $\alpha$-RuCl$_3$ is not known, as the 0.3-8.3 meV energy scale is well below the range of expected optical transitions. Excitations across the Mott gap onset at 200 meV ($\sim 50$ THz)~\cite{plumb2014alpha} and the dominant dipole-active optic phonon resonance is found at $\sim$ 35 meV ($\sim$ 8.5 THz) (see SI section \upperRomannumeral{5} \cite{SI}).  Lorentzian fits to this phonon mode yield $\sigma_1(\omega)$ that is well below the measured value near 1 meV (see SI section \upperRomannumeral{5} \cite{SI}). Although non-Lorentzian lineshapes are found in many wide-gap insulators, the signatures of these acoustic-phonon assisted processes are strong temperature and featureless power-law frequency dependences~, both of which are inconsistent with the spectra of Fig. 3.

Given the structure in the spectra on the meV energy scale, we believe it is possible that the THz absorption is related in some way to the spin-degree of freedom. We note that features of the THz spectra, particularly the linear in $\omega$ low-energy cut-off shown in Fig. 1a and 2c, closely resemble the dynamical spin structure factor predicted for the Kitaev-Heisenberg Hamiltonian \cite{song2016low}. Intrinsic mechanisms by which spin fluctuations in frustrated magnets acquire electric-dipole activity were described in Refs. \cite{bulaevskii2008electronic} and \cite{potter2013mechanisms}. The predicted optical conductance per atomic layer is either $\sim(e^2/h)(t/U)^3$ if the lattice is fixed, and $\sim(e^2/h)(t/U)^2$ if magnetoelastic coupling is considered ($t$ and $U$ are hopping and Coulomb energies, respectively). Converting the spectra shown in Fig. 3 to conductance per Ru layer (see SI section VI) yields an optical conductance of $\sim 10^{-4}(e^2/h)$ and of the same order as found in the Kagome compound Herbertsmithite \cite{pilon2013spin}.

To summarize, we have measured the optical absorption of $\alpha$-RuCl$_3$ at photon energies comparable to its magnetic exchange interactions, revealing a sharp magnon resonance and broad-band optical conductivity that cuts-off linearly below 1 meV.  We tracked the evolution of the frequency, damping rate, and spectral weight of the dynamic susceptibility of the $\textbf{q}=0$ magnon as a function of magnetic field.  We believe this information is critical to understanding the role of the Kitaev and other, ``parasitic", exchange interactions in determining the nature of the quantum critical points and novel phases of $\alpha$-RuCl$_3$.

We thank N. P. Armitage, L. Balents, C. Batista, A. Potter, M. Serbyn, S. Winter, R. Valenti and A. Vishwanath for helpful discussions, and H. Bechtel and M. Martin for support at the Advanced Light Source beamline 1.4.3 and 1.4.4. Terahertz spectroscopy was performed at Lawrence Berkeley National Laboratory in the Spin Physics program supported by the Director, Office of Science, Office of Basic Energy Sciences, Materials Sciences and Engineering Division, of the U.S. Department of Energy under Contract No. DE-AC02-76SF00515. A.L. and L.W. were supported by the Gordon and Betty Moore Foundation's EPiQS Initiative through Grant GBMF4537 to J.O. at UC Berkeley. The work at ORNL was supported by the US-DOE, Office of Science, Basic Energy Sciences, Materials Sciences and Engineering Division, under contract number DE-AC05-00OR22725. A.B. and S.N. were supported by the US DOE, Office of Science, Office of Basic Energy Sciences, Division of Scientific User Facilities. P. L. K. and D. M. acknowledge support from Gordon and Betty Moore Foundation's EPiQS Initiative through Grant GBMF4416. 

A. L. and L. W. contributed equally to this work.

\bibliography{RuCl}

\begin{thebibliography}{58}
\bibitem[{not()}]{note_structure}
\bibinfo{note}{Note that trigonal/rhombohedral structure has 3-fold rotation
  symmetry along the stacking direction (c) of RuCl$_{3}$.}
\end{thebibliography}

\pagebreak
\widetext
\begin{center}
\textbf{\large Supplementary Information for ``Antiferromagnetic resonance and terahertz continuum in $\alpha-$RuCl$_3$''}
\end{center}

\setcounter{equation}{0}
\setcounter{figure}{0}
\setcounter{table}{0}
\setcounter{page}{1}
\makeatletter
\renewcommand{\theequation}{S\arabic{equation}}
\renewcommand{\thefigure}{S\arabic{figure}}

\title{
Supplementary Information for ``Antiferromagnetic resonance and terahertz continuum in $\alpha-$RuCl$_3$''}
\maketitle


\section{Time-domain TH\lowercase{z} spectroscopy}
\subsection{The spectrometer system}
THz pulses were generated by an Auston switch consisting of a dipolar electrode antenna patterned onto a semiconductor. An ac bias voltage is applied across the electrodes and an ultrafast 100 fs laser pulse excites free carriers that are accelerated by the bias voltage, emitting THz radiation. The THz pulses are focused onto the sample by off-axis parabolic mirrors and the transmitted radiation is collected by a receiver antenna. Two grid-patterned polyethylene polarizers were used to perform polarized transmission measurements. The first polarizer was placed directly before the sample (but outside of the cryostat) and was set parallel to either crystal axis $a$ or $b$. The second polarizer was placed directly outside of the cryostat after the sample and was oriented parallel to the direction of polarization of the Auston switch detector. For field dependent measurements, the DC magnetic field is applied by a 7 Tesla split-coil superconducting magnet in the ab plane of the sample.

\subsection{Analysis of transmission coefficient}

In time-domain spectroscopy, weak absorption leads to a sequence of transmitted pulses in which the first traverses the sample once and subsequent ``echo" pulses undergo multiple reflections and traversals. For the measurements reported in this work, the sample is sufficiently thick ($\sim$0.8mm) and the round trip time sufficiently long that the echoes are well-separated in time. In this regime, we can restrict our time delay window to the first pulse. The transmission coefficient for a single pass through the sample is given by
\begin{equation}
t(\omega)=[4\tilde{n}/(\tilde{n}+1)^2]\exp{[ik_0(\tilde{n}-1)d]},
\end{equation}
where $\tilde{n}=n+i\kappa$ is the complex index of refraction, $k_0$ is the free space wavevector, and $d$ is the sample thickness. In $\alpha$-RuCl$_3$ at THz frequencies we have $n\gg\kappa$, in which case,
\begin{equation}
|t(\omega)|\cong[4n/(n+1)^2]\exp{[-\alpha(\omega)d]},
\end{equation}
where the absorption coefficient $\alpha(\omega)$ is given by $\omega\kappa(\omega)/c$.

\subsection{Conversion of absorption coefficient to response functions}

The propagation of THz through media is described by Maxwell's equations in matter, together with the standard constitutive relations,
\begin{equation}
\bf{D}(\omega)=\epsilon(\omega)\bf{E}(\omega)
\end{equation}
and
\begin{equation}
\bf{B}(\omega)=\mu(\omega)\bf{H}(\omega),
\end{equation}
where $\epsilon(\omega)$ is the dielectric function and $\mu(\omega)$ is the magnetic permeability, related to the magnetic susceptibility, $\chi(\omega)$ by $\mu(\omega)=\mu_0[1+\chi(\omega)]$. Together these relations yield,
\begin{equation}
\tilde{n}(\omega)=\sqrt{\frac{\epsilon(\omega)\mu(\omega)}{\epsilon_0\mu_0}}.
\end{equation}
In the THz region of the $\alpha$-RuCl$_3$ spectrum, we have the strong inequalities, $\mu_2/\mu_0,\epsilon_2/\epsilon_0\ll\epsilon_1/\epsilon_0$, where the subscripts 1, and 2 denote real and imaginary parts, respectively. To a very good approximation, we can express the imaginary part of the index of refraction, $\kappa(\omega)$, as the sum of contributions from the dielectric and magnetic response functions,
\begin{equation}
\kappa(\omega)\cong\frac{n}{2}\left(\frac{\epsilon_2(\omega)}{\epsilon_1}+\chi_2(\omega)\right).
\end{equation}
The absorption coefficient, $\alpha(\omega)$, can then be written in the form used in the main text,
\begin{equation}
\alpha(\omega)\cong\frac{n\omega\chi_2(\omega)}{2c}+\frac{\sigma_1(\omega)Z_0}{2n},
\end{equation}
where the real part of the optical conductivity $\sigma_1(\omega)=\omega\epsilon_2(\omega)$, and $Z_0\equiv\sqrt{\mu_0/\epsilon_0}$ is the impedance of free space. Thus, in this regime the absorption is the sum of contributions from the magnetic and dielectric response functions.

The factor, $nd/c$, is evaluated directly from the time delay between the arrival of the first and second pulses at the detector. This time interval is the roundtrip time, $T_{rt}$, which is equal to $2nd/c$. As stated in the main text, determination of $\chi_2(\omega)$ from the magnetic-dipole component of the absorption is self-calibrated, as it is extracted using only observables obtained from our measurement, that is,
\begin{equation}
\chi_2(\omega)=\frac{4\alpha(\omega)d}{\omega T_{rt}}.
\end{equation}

\section{Registration of principal optical axes and crystal axes}

\begin{figure}[h]
   \centering
    \includegraphics[width=0.6\linewidth]{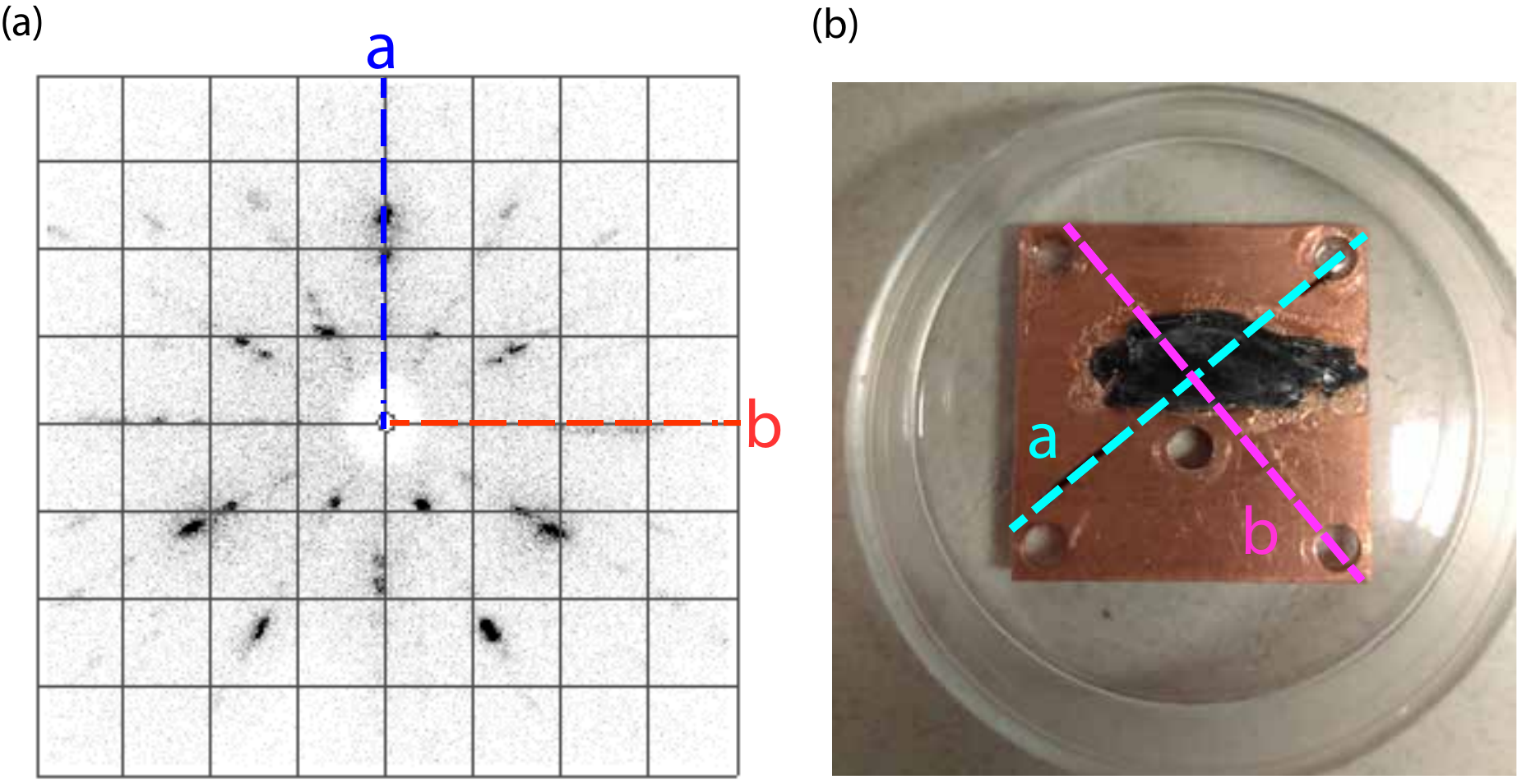}
    \\
    FIG S1: (a) Laue diffraction pattern showing reciprocal of hexagonal lattice. The optic $a$ and $b$ axes are denoted by dashed lines. (b) The orientation of $a$ and $b$ axes on the sample.
 \end{figure}
We performed Laue X-ray diffraction in order to register the principal optic axes observed in the THz measurement with the crystal axes. With the sample in the same mount as used in THz measurements, we aligned the sample in a Laue diffractometer such that the optic axes as determined by THz were oriented vertically and horizontally.  The resulting Laue diffraction pattern is shown in Fig. S2 (a), and the reciprocal lattice of the hexagonal structure is apparent. We find that the principal axes observed in the THz measurement indeed correspond to the symmetry axes of the crystal. The alignment of this pattern with the optic axes allows us to determine that the direction of the terahertz magnetic field showing the strongest antiferromagnetic resonance corresponds to the  $b$ axis, as shown in Fig. 1 of the main text.

\section{Determination of Index of Refraction}

\begin{figure}[h]
   \centering
    \includegraphics[width=0.6\linewidth]{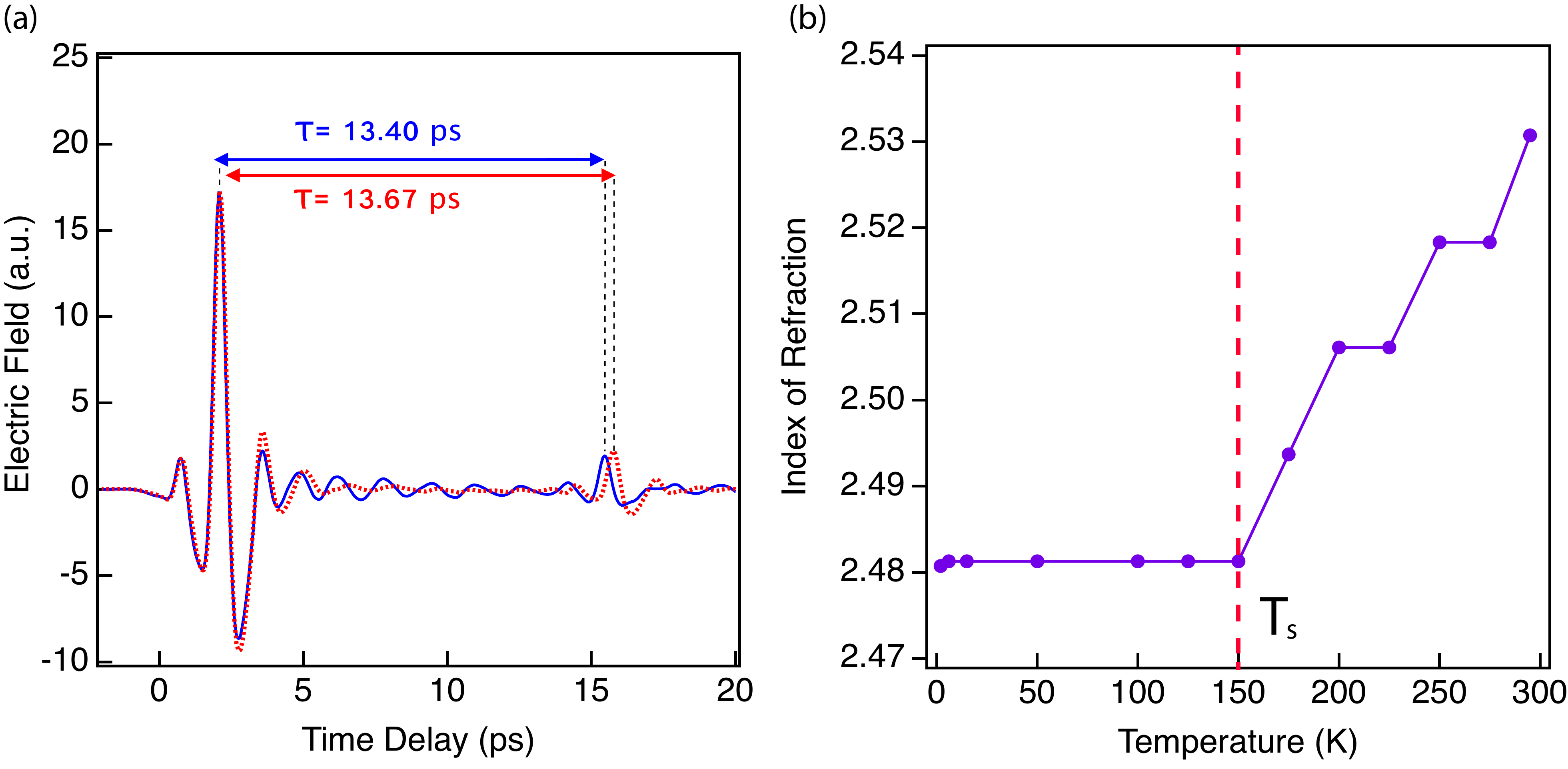}
    \centering
    \\
    FIG S2: (a) THz transient including the initial transmitted pulse and the echo pulse at 4 K (blue) and 294 K (red). Arrows indicate the 0.27 ps shift of the echo pulse between the two temperatures. (b) Index of refraction calculated from echo time as a function of temperature.
 \end{figure}
In order to determine the index of refraction of our samples of $\alpha$-RuCl$_{3}$, we measured the time delay between the initial transmitted THz pulse and subsequent ``echo" pulse. The initial pulse makes a single pass through the sample, while the echo pulse undergoes two internal reflections and makes three passes through the sample. The delay time between the initial pulse and the first echo is given by $T_{rt} = 2nd/c$, where $n$ is the index of refraction, $d$ is the sample thickness, and $c$ is the speed of light. In determining the index, we used a sample with a large uniform thickness, measured by digital micrometer to be 0.810$\pm$0.015 mm.

Fig. S2 (a) shows initial and echo THz pulses at 4 K and 295 K sampled at time delay intervals $\Delta t= 0.667$ picoseconds. At 4 K, the delay time between the peak of the initial and echo pulse is 13.40 ps, rounding to four significant digits, which corresponds to $n$ = 2.48. The comparison of the two pulse sequences in Fig. S2(a) shows that $n$ depends weakly on $T$. As shown in Fig. 1(b), $n$ is $T$ independent at low $T$ but increases linearly with $T$ starting at 150 K, reaching 2.53 at 294 K. The measurement was taken upon warming the sample,  and the observed jump in the index around 150 K is consistent with the structural phase transition reported in Ref. \cite{ziatdinov2016atomic}. For the analysis in the main text, the rounded value $n$ = 2.5 was used to evaluate  $\sigma_1$ at all temperatures.

\section{Spin wave analysis}

\subsection{Zero magnetic field}

\begin{figure}[h]
 \centering
    \includegraphics[width=\linewidth]{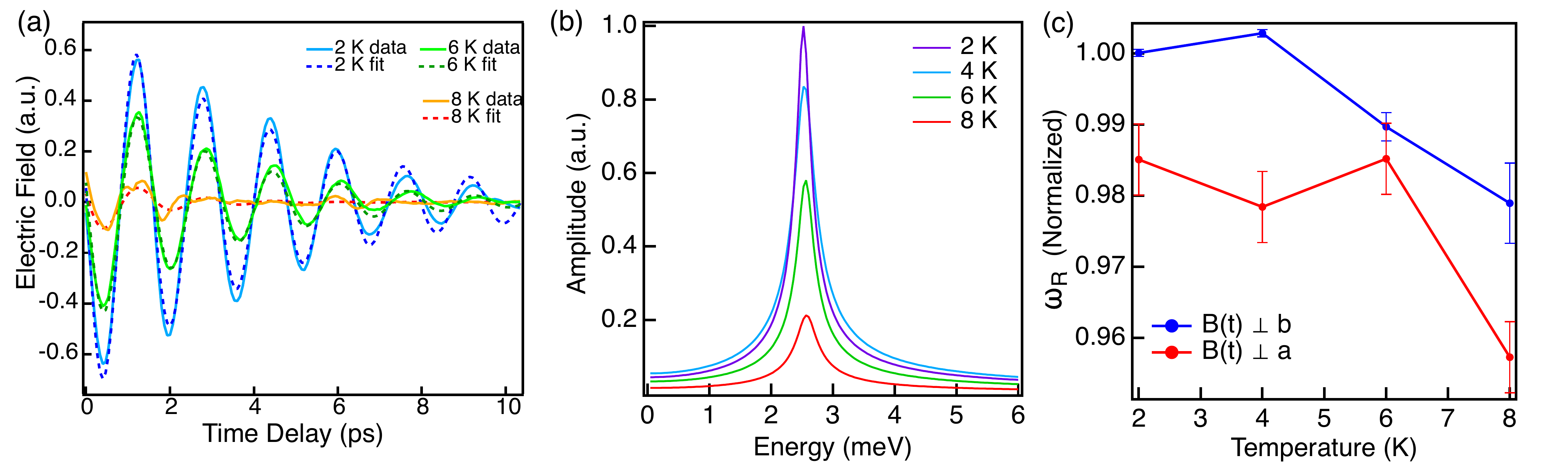}
     \centering
\\
    FIG. S3:  (a) Solid lines show time traces at 2, 4, 6, and and 8 K after subtracting the time trace at 15 K. Fits to exponentially damped sinusoidal oscillations are shown as dashed lines. (b) Lorentzian peaks extracted from fits shown in (a). (c) AMFR center frequency for axes a and b, as a function of temperature.
 \end{figure}

To determine the amplitude, width, and energy of the AFMR as a function of temperature, we first subtracted the pulse measured at 15 K from pulses measured below the N\'eel temperature. The result of the subtraction is shown in Fig. 2 of the main text and reproduced in Fig. S3 (a) here. The fitting function is an exponentially damped sine wave $A\exp{(-\Gamma t)}\sin(\omega_{R}t)$ where $A$, $\Gamma$, and $\omega_{R}$ are fit parameters. In Fig. S3, the fits (dashed lines) are shown in comparison to the raw data (solid lines) for three temperatures below $T_N$. From the fit parameters, we extract the temperature dependence of the amplitude and resonant frequency of the AMFR as the sample is warmed through $T_{N}$ as shown in Fig. S3 (b) and (c). Approaching $T_{N}$, we observe a small redshift in the resonant frequency, as observed in classical collinear antiferromagnets CoCl$_2$ \cite{jacobs1965antiferromagnetic}.

\begin{figure}[h]
 \centering
    \includegraphics[width=0.4\linewidth]{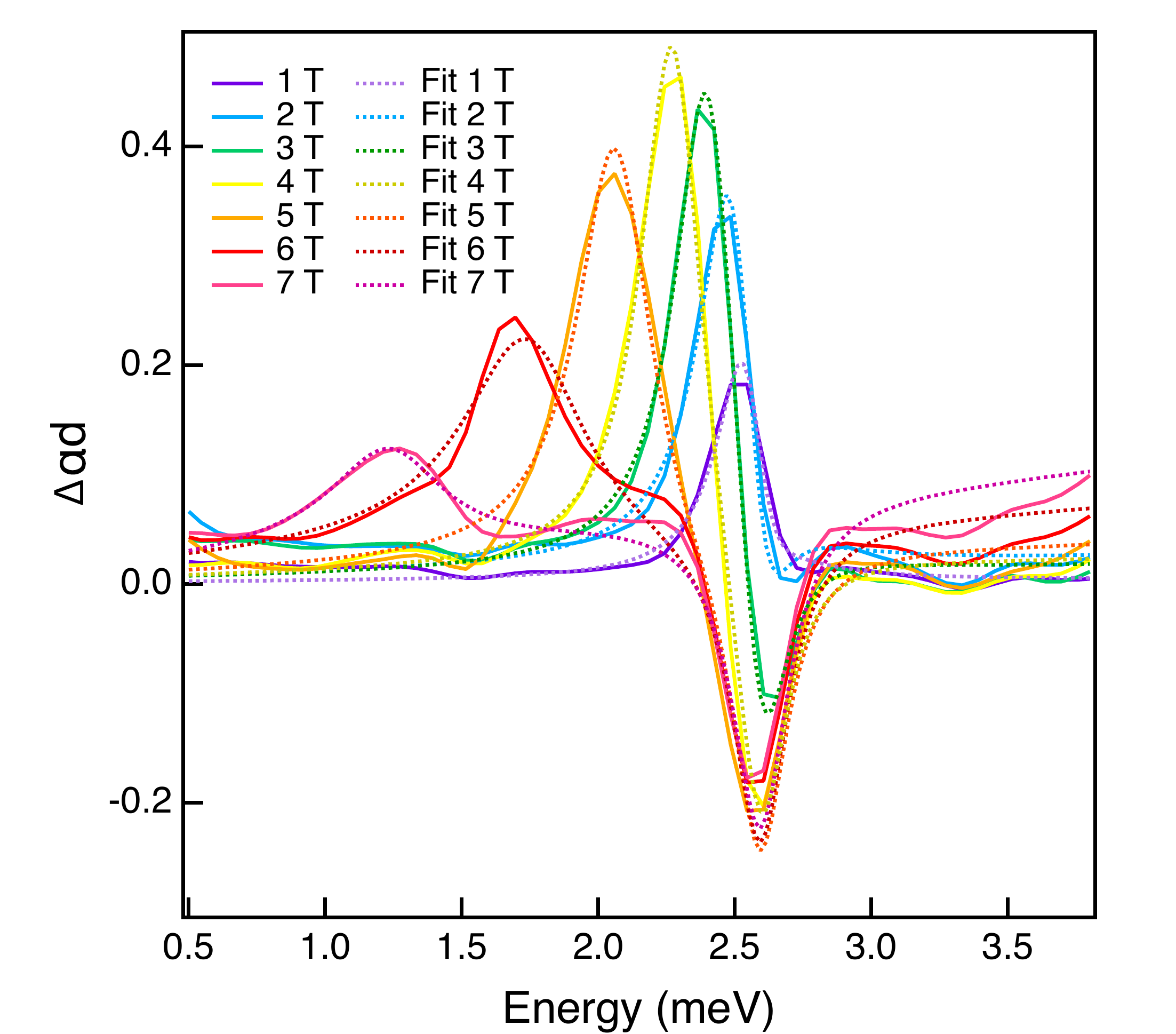}
     \centering
\\
   FIG. S4: Absorption difference spectra with the $B=$0 T spectrum subtracted data (solid lines) and fits (dashed lines) at magnetic fields 1 T through 7 T and temperature $T=$ 4 K.
\end{figure}

\subsection{AFMR vs. magnetic field}

To analyze the AFMR as a function of magnetic field, we subtract the absorption spectrum at 0 Tesla from the spectra at 1 - 7 Tesla and fit to a Lorenztian lineshape:
\begin{equation}
\Delta\alpha(\omega,B)d=\frac{T_{rt}S_{1}\omega^2\Gamma_{1}}{4((\omega^2-\omega_{R1}^2)^2+\omega^2\Gamma_1^2)}-\frac{T_{rt}S_{0}\omega^2\Gamma_{0}}{4((\omega^2-\omega_{R0}^2)^2+\omega^2\Gamma_0^2)} +c\omega,
\end{equation}
where $S_0$, $\Gamma_0$, and $\omega_{R0}$ are the parameters for the 0 T spectrum, $T_{rt}$ is the pulse round trip time, and $c$ is the slope of a linear in $\omega$ background. For the fits and data shown in Fig. S4, the DC magnetic field $B$ is parallel to the THz magnetic field, \textbf{B(t)}, which is at an angle of 45 degrees from the $a$ and $b$ principal axes of the crystal. Note that when we obtain the absorption difference spectra in Fig 2 (d) and Fig. S4, we scan longer in the time domain to increase the frequency resolution. Fabry-Perot oscillations that result from including the first echo are greatly reduced in the difference spectra. However, for the absorption curves plotted in Fig. 2c, we used time-windowing to avoid  Fabry-Perot oscillations. As a result, the AMFR is slightly broadened relative to the difference spectra.

In $\alpha$-RuCl$_3$ we find that AFMR shifts to lower frequency in an applied magnetic field, suggesting that $B$ acts as a knob that tunes the system towards a quantum phase transition. This is quite different from what is seen in conventional collinear antiferromagnets where the zero-field AFMR generally splits and shifts as expected from the Zeeman interaction \cite{keffer1952theory, ross2015antiferromagentic}. Notably, the field dependence we observe differs as well from what is seen in other frustrated magnets such as LiZn$_2$Mo$_3$O$_8$~\cite{sheckelton2014local}, Yb$_2$Ti$_2$O$_7$~\cite{pan2014low}, Cs$_2$CuBr$_4$ and Cs$_2$CuCl$_4$~\cite{zvyagin2014direct}, where the energy of the modes increases with the applied field. This review of the literature suggests that our study may be the first investigation of the frequency, damping, and spectral weight of the zero-wavevector magnon as a system approaches the quantum phase transition from an antiferromagnetic to spin-disordered state.

\section{Fourier Transform Infrared spectra}
To quantify the contribution to the low-frequency conductivity from optically-active phonons, we performed Fourier transform infrared spectroscopy (FTIR) spectroscopy at room temperature on a sample with dimensions 6 mm $\times$ 1 cm and order of 100 $\mu$m thickness. The sample was mounted on a 5 mm aperture and the transmission was measured at room temperature for unpolarized infrared (IR) radiation with energy range of $\sim$100 - 10,000 cm$^{-1}$ or $\sim$ 3 - 330 THz. The main features are an absorption resonance at 5 THz and a (Restrahl) band of zero transmission from 8-10 THz, which results from near unity reflectivity in this range. This spectrum is consistent with earlier studies of reflectivity of $\alpha$-RuCl$_3$ \cite{guizzetti1979fundamental}.

\begin{figure}[h]
        \centering
        \includegraphics[width=0.85\linewidth]{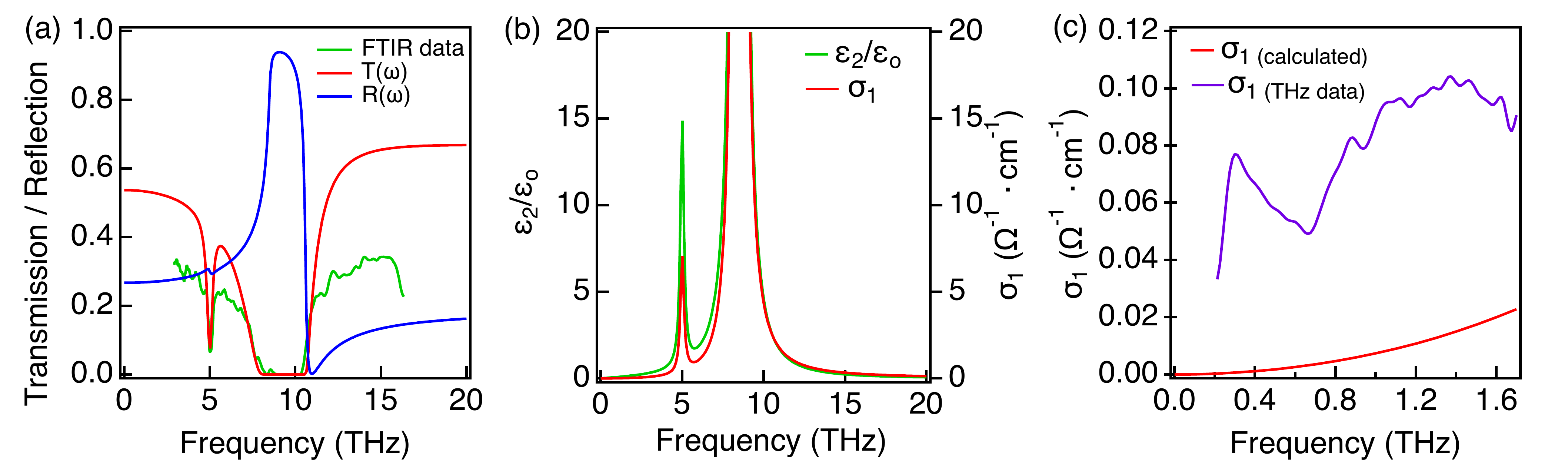}
        \\
       FIG. S5: (a) FTIR transmission data (green) compared with theoretical transmission (red) and reflection (blue) as modeled by Eq. S1. (b) The dimensionless imaginary dielectric constant, $\epsilon_{2}/\epsilon_{0}$ (green) as extracted from our model and the associated optical conductivity, $\sigma_1$ (red). (c) Measured THz conductivity at 294 K (purple) compared with the calculated phonon contribution to the optical conductivity (red).
\end{figure}

We can model the contribution of the two low-energy phonons to the transmission spectrum using a Lorentzian fitting function of the form,
\begin{equation}
\epsilon(\nu)=\epsilon_\infty+\frac{\Omega_1^2\epsilon_0}{\nu_1^2-\nu^2-i\nu\gamma_1}+\frac{\Omega_2^2\epsilon_0}{\nu_2^2-\nu^2-i\nu\gamma_2}.
\end{equation}
The factors $\Omega_i$, $\nu_i$, and $\gamma_i$, where $i=1,2$ are the spectral weight, resonant frequency, and damping parameters, respectively for the two phonon modes. Fig. S5 (a) shows the transmission, $T(\omega)$, and reflectivity, $R(\omega)$ calculated using parameters $\Omega_1=16$ THz, $\nu_1=8.5$ THz, and $\gamma_1=12$ GHz to fit the Restrahl feature, and $\Omega_2=1.25$ THz, $\nu_1=5.0$ THz, and $\gamma_1=0.25$ THz to describe the absorption peak. The spectral weight of the phonon that produces the Restrahl feature is a factor of order $10^2$ larger than that of the lower energy peak.  The steepness, bandwidth and near null transmission of the Restrahl band constrain its spectral weight parameter within an error range of about 10$\%$.

By fitting the phonon features in the transmission spectra, we can estimate the optical conductivity at the lower end of our spectral range, 0.2 THz, associated with the Lorentzian tail of the optical phonon absorption. Fig. S5 (b) shows the calculated $\epsilon_2/\epsilon_0$ on the left hand scale and the corresponding $\sigma_1$ on an expanded right hand scale. With the spectral weight and damping parameters quoted above, the imaginary part of the dimensionless dielectric function at 0.2 THz is $\epsilon_2/\epsilon_0=2\times 10^{-3}$. This value for the imaginary part of the dielectric function corresponds to $\sigma_1=2\times10^{-4}$ $\Omega^{-1}\cdot$ cm$^{-1}$, which is approximately two orders of magnitude smaller than the optical conductivity in the THz regime (the comparison is illustrated in Fig. S5 (c)). Finally, we note that fact that the THz absorption depends only weakly on $T$ argues against effects such as multi-phonon absorption or defect-activated conductivity, which are typically highly dependent on $T$, as illustrated by references \cite{sparks1982simple} and \cite{helgren2004frequency}.

\section{Conversion of optical conductivity to conductance per R\lowercase{u} plane}

\begin{figure}[h!]
 \centering
    \includegraphics[width=0.6\linewidth]{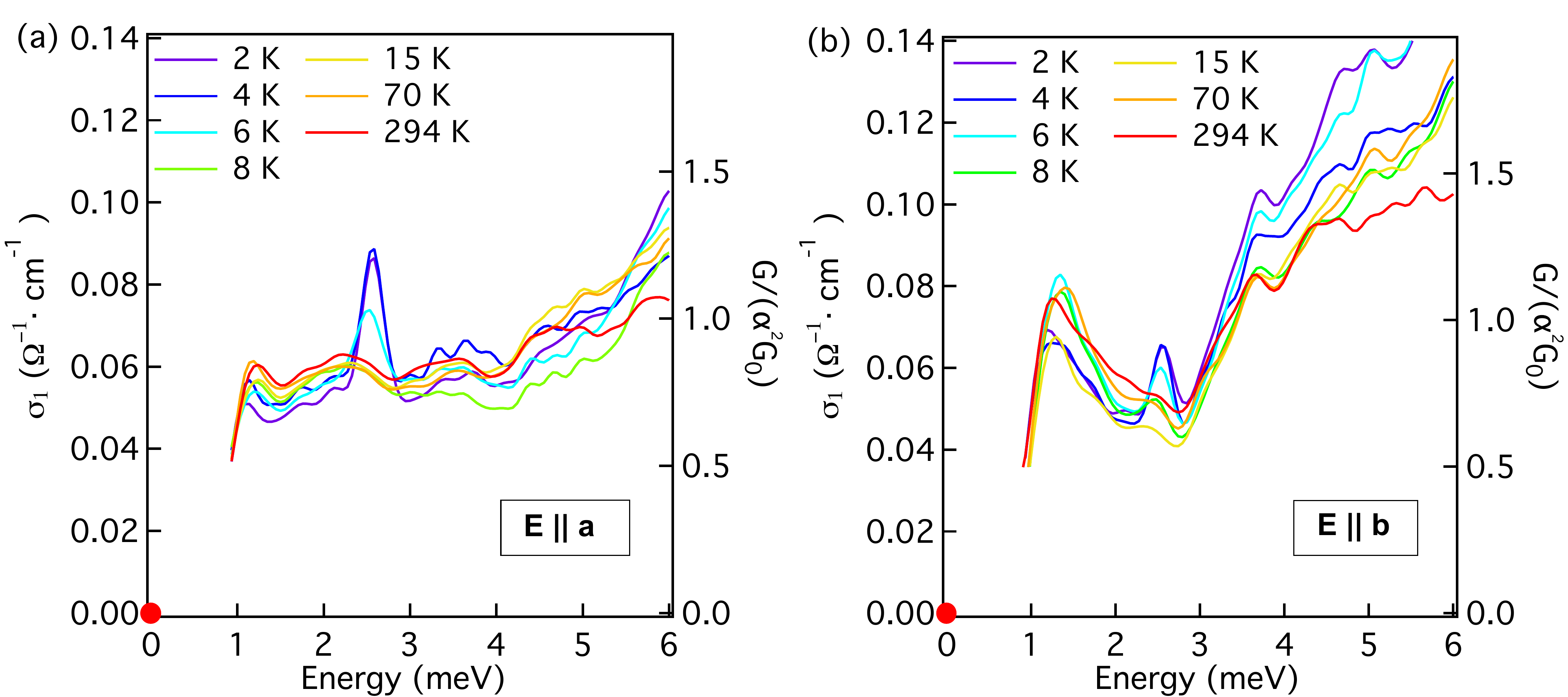}
	\\
	Fig. S6: (a) Absorption spectra interpreted as optical conductivity, with E parallel to a. (b) Same for axis b. Right-hand scales show equivalent conductance per Ru plane normalized by the product of the quantum of conductance and the fine structure constant squared.
\end{figure}

Although the absorption continuum is too strong to arise from direct coupling to magnetic dipole moments, it is weak compared to fully allowed electric dipole transitions. To illustrate this, we convert $\sigma_1(\omega)$ to a dimensionless measure of dissipation strength -- the conductance per Ru honeycomb plane normalized by $\alpha_f^2 G_0$, where $\alpha_f$ is the fine structure constant and $G_0=e^2/h$ is the quantum of conductance. The scale on the right-hand axes of Fig. S6 illustrates the results of this conversion.  Although it is unclear at present if $\alpha_{f}$ is relevant to optical conductivity in $\alpha$-RuCl$_3$, the dimensionless parameter emphasizes that $G(\omega)$ per honeycomb plane is extremely small, that is, reduced from $G(\omega)$ of a weakly interacting hexagonal system such as graphene by a factor of roughly $2\times 10^4$ \cite{novoselov2005two}.

\section{Measurement of DC resistivity}
\begin{figure}[h!]
        \centering
        \includegraphics[width=0.35\linewidth]{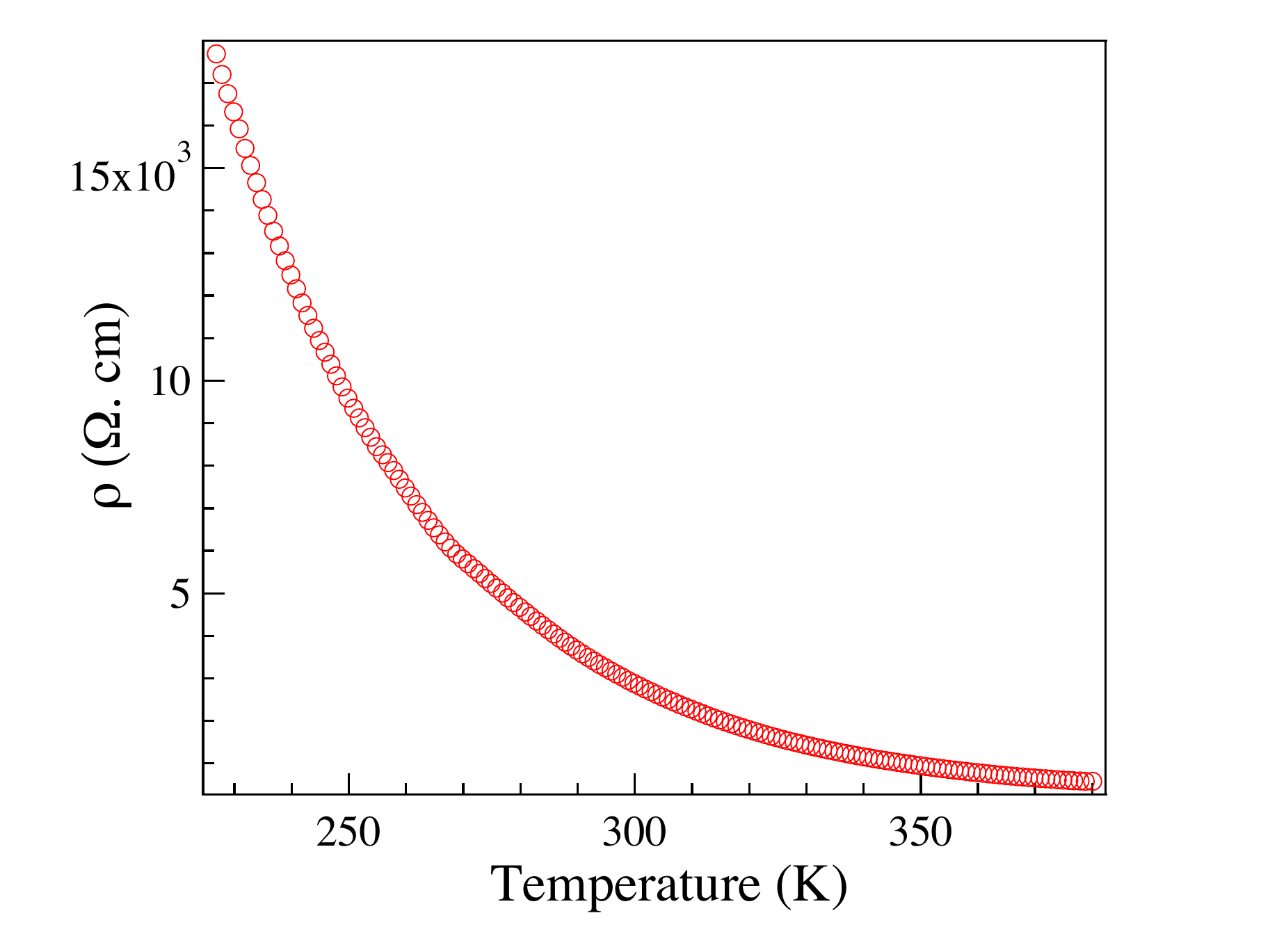}
        \\
       FIG. S7: $dc$ resistivity, $\rho$ as a function of temperature.
\end{figure}
The dc resistivity of $\alpha$-RuCl$_{3}$ was measured  for temperatures above 200 K as shown in Fig. S7. We note that the resistivity measured at 300 K is $\rho \sim 3 \times 10^{-3}$ $\Omega \cdot$ cm, corresponding to a conductivity $\sigma_{dc}=3 \times 10^{-4}$ $\Omega^{-1}\cdot$ cm$^{-1}$. This value is indistinguishable from the origin of the plots shown in Fig. S7.

\end{document}